\documentclass[aps,pra,twocolumn,floatfix,showpacs,citeautoscript,superscriptaddress]{revtex4-1}
\usepackage{graphicx}
\usepackage{amsmath}
\usepackage{amsfonts}
\usepackage{mathrsfs}
\usepackage{color}

\usepackage{amssymb}
\usepackage{bm}
\usepackage{times}
\usepackage{MnSymbol}
\usepackage{lipsum}

\begin{document}

\title{Fate of dynamical phases of a BCS superconductor beyond the dissipationless regimen}

\author{H. P. {Ojeda Collado}} 
%\affiliation{Centro At\'omico Bariloche and Instituto Balseiro,
%  Comisi\'on Nacional de Energ\'ia At\'omica, 8400 Bariloche,
%  Argentina}
\affiliation{Centro At{\'{o}}mico Bariloche,
Comisi\'on Nacional de Energ\'{\i}a At\'omica, 8400 Bariloche, Argentina}
\affiliation{Instituto Balseiro, Universidad Nacional de Cuyo, 8400 Bariloche, Argentina}
\affiliation{Consejo Nacional de Investigaciones Cient\'{\i}ficas y T\'ecnicas (CONICET), Argentina}
\author{Gonzalo Usaj}
\affiliation{Centro At{\'{o}}mico Bariloche,
Comisi\'on Nacional de Energ\'{\i}a At\'omica, 8400 Bariloche, Argentina}
\affiliation{Instituto Balseiro, Universidad Nacional de Cuyo, 8400 Bariloche, Argentina}
\affiliation{Consejo Nacional de Investigaciones Cient\'{\i}ficas y T\'ecnicas (CONICET), Argentina}\author{Jos\'{e} Lorenzana}
%\email[Corresponding author: ]{jose.lorenzana@cnr.it}
\affiliation{ISC-CNR and Department of Physics, Sapienza University of Rome, Piazzale Aldo Moro 2, I-00185, Rome, Italy}
\author{C. A. Balseiro}
%\email[Corresponding author: ]{balseiro@cab.cnea.gov.ar}
\affiliation{Centro At{\'{o}}mico Bariloche,
Comisi\'on Nacional de Energ\'{\i}a At\'omica, 8400 Bariloche, Argentina}
\affiliation{Instituto Balseiro, Universidad Nacional de Cuyo, 8400 Bariloche, Argentina}
\affiliation{Consejo Nacional de Investigaciones Cient\'{\i}ficas y T\'ecnicas (CONICET), Argentina}

\begin{abstract}
The BCS model of an isolated superconductor initially prepared in a nonequilibrium state, predicts the existence of interesting dynamical phenomena in the 
time-dependent order parameter as decaying oscillations, persistent oscillations and overdamped dynamics. To make contact with real systems remains an open 
challenge as one needs to introduce dissipation due to the environment in a self-consistent computation. Here, we reach this goal with the use of 
the Keldysh formalism to treat the effect of a thermal bath. We show that, contrary to the dissipationless case, all dynamical phases reach the equilibrium order 
parameter in a characteristic time that depends on the coupling with the bath.  Remarkably, as time evolves, the overdamped phase shows a fast crossover where the 
superconducting order parameter recovers  to  reach a state with a well-developed long range order that tends towards equilibrium with the damped Higgs mode 
oscillations. Our results provide a benchmark for the description of the dynamics of real out-of-equilibrium superconductors relevant for quantum technological 
applications. 
\end{abstract}
% insert suggested PACS numbers in braces on next line
% insert suggested keywords - APS authors don't need to do this
%\keywords{}
\date{\today}
\maketitle

%%%%%%%%%%%%%%%%%%%%%%%%%%%%%%%%%%%%%%%
\section{Introduction}
%%%%%%%%%%%%%%%%%%%%%%%%%%%%%%%%%%%%%%%
The recent advances in experimental pump-probe techniques offer new opportunities to study out-of-equilibrium states of matter with 
collective modes or phases that are not accessible with more conventional tools \cite{Stojchevska2014,Mitrano2016,Fausti2011}. A notable example include the 
observation of oscillations of the condensate in superconductors with a frequency determined by the superconducting gap \cite{Mansart2013,Matsunaga2013,Matsunaga2014}.  
However, the study of out of equilibrium interacting systems presents a demanding 
challenge. From the theoretical point of view, the problem requires the precise implementation of the Baym-Kadanoff-Keldysh non-equilibrium quantum field theory \cite{Stefanucci,Stan2009}, a 
strategy that although well formulated it is sometimes difficult to realize in practice. Nevertheless, some particular cases have been studied with a good degree of 
control \cite{Stefanucci2009,Leeuwen2007}. Diagrammatic expansions and dynamical mean-field theories for out of equilibrium fermions on a lattice are examples of the cutting-edge developments \cite{Ecksteini2009,Ecksteini2014}.

The cases of superconductivity on fermionic condensates of cold atoms are exceptional and have been studied by several groups during the last years \cite{Barankov2004,Barankov2006a,Yuzbashyan2006a,Hannibal2018,Hannibal2018a}. The integrability of the reduced BCS Hamiltonian in the dissipationless regime allows for a precise formulation of the time-dependent out of equilibrium dynamics when 
the microscopic parameters change with time \cite{Yuzbashyan2005,Yuzbashyan2006}. It has been shown that after a sudden change of the pairing interaction $\lambda$ the system evolves towards three 
distinct stationary dynamical phases \cite{Barankov2006a,Yuzbashyan2006}. After a small change of $\lambda$, either an increase or a decrease, the order parameter shows oscillations of frequency 
$2\Delta_{\infty}$ and a power law decay ($t^{-1/2}$) reaching a long time asymptotic value $\Delta_{\infty}$. The decay of the oscillations is due to the dephasing of the excitations and this regime is known as the dephasing phase (phase I). 
The asymptotic value $\Delta_{\infty}$ is always smaller than the corresponding thermodynamical
equilibrium value due to the proliferation of out of equilibrium pair excitations. If the coupling is reduced by a large amount, the dynamics
becomes overdamped and the asymptotic value of the long-range order parameter becomes zero (phase II). On the other hand, if the coupling increases above a critical value, 
the order parameter shows persistent oscillations as quasiparticles evolve synchronously driven by the self-consistent pairing field (phase III).

The case of a periodically driven or pumped superconductor is also very interesting. The order parameter shows a synchronization phenomena of Rabi oscillations of 
quasiparticles states which can be exploit to access all the aforementioned dynamical phases \cite{HP2018}. 

Notwithstanding these interesting theoretical findings, the convergence of theory and experiment in this field is quite problematic.
One the one hand, the integrability of the BCS model implies that an out of equilibrium system  can not reach thermal equilibrium, no matter how long the system is 
allowed to evolve. One the other hand, real systems do of course relax and if the relaxation is too fast, $2\Delta_{\infty}$-like oscillations
will not be visible. Fortunately, pump-probe experiments performed in cuprates~\cite{Mansart2013}
and in Nb$_{1-x}$Ti$_{x}$N \cite{Matsunaga2013,Matsunaga2014} show that $2\Delta_{\infty}$-like oscillations are visible. 
Therefore, relaxation times are long enough to have access to a regime where energy-conserving out of equilibrium dynamics dominates the system response. 

It remains the theoretical challenge to describe the interesting crossover from the out of equilibrium regime to a thermal state. An obvious choice is to 
consider a thermal bath which can exchange energy with the superconductor.
The theoretical formulation of such nonequilibrium many-body problem with dissipation poses a demanding issue. Indeed, to consider explicitly all the degrees of freedom of a very large bath is numerically unaffordable. For a finite thermal bath spurious oscillations appear and the non-thermalization problem is not solved.

In order to formulate the problem in a feasible and computationally accessible way, different approximation schemes have been proposed in the context of pumped 
s-wave superconductors in which the order parameter varies with time. Recent studies, of the time-resolved angle-resolved photoemission spectroscopy (tr-ARPES) \cite{Moore2019} and 
the optical conductivity \cite{Millis2017}, have adopted similar approaches. Namely,  the  inclusion of inelastic scattering processes that releases the extra energy via a self-energy in the 
lesser Green function leading to properly recovery of thermalization at long times. However, none of these works computed the self-consistent dynamics of the order parameter. In these cases, that simulate the effect of light-field pump pulses, the order parameter was taken as a known function of time that drives the system out of equilibrium breaking the time invariance and 
strongly modifying the system response. 

In this work we present a {\em self-consistent} calculation of an out of equilibrium s-wave superconductor including relaxation due to the coupling of the system 
to an external bath. We illustrate the method by considering a BCS superconductor and the simplest non-equilibrium protocol of a quantum quench of the interaction parameter.
We show that at short or moderate times after the quench, traces of the Barankov-Levitov dynamical phase diagram (see Ref. \cite{Barankov2006a}) are clearly observed while at long times new equilibrium states are recovered.

%%%%%%%%%%%%%%%%%%%%%%%%%%%%%%%%%%%%%%%%%%%%%%%%%%%%%%%%%%%
\section{The BCS Hamiltonian and the problem formulation}
%%%%%%%%%%%%%%%%%%%%%%%%%%%%%%%%%%%%%%%%%%%%%%%%%%%%%%%%%%%

We consider a single-band s-wave superconductor described by the Hamiltonian
%%%%%%%%%%%%%%%%%%%%%%%%%%%%
\begin{equation}
\label{eq:HBCS}
H_{\mathrm{BCS}}=\sum_{\bm{k},\sigma}\xi_{\bm{k}}c_{\bm{k}\sigma}^{\dagger}c_{\bm{k}\sigma}^{}-\lambda(t)\sum_{\bm{k},\bm{k^{\prime}}}c_{\bm{k}\uparrow}^{\dagger}c_{\bm{-k}\downarrow}^{\dagger}c_{\bm{-k^{\prime}}\downarrow}^{}c_{\bm{k^{\prime}}\uparrow}^{}
\end{equation}
%%%%%%%%%%%%%%%%%%%%%%%%%%%%
where  $c_{\bm{k}\sigma}$ ($c_{\bm{k}\sigma}^{\dagger})$
destroys (creates) an electron with momentum $\bm{k}$, energy $\varepsilon_{\bm{k}}$ and spin
$\sigma$. Here  $\xi_{\bm{k}}=\varepsilon_{\bm{k}}-\mu$ measures the energy from the Fermi level $\mu$.
The pairing interaction $\lambda(t)$ is allowed to be time-dependent. We will focus on 
the effect of a reservoir on the dynamical phases that are obtained after a quench of 
$\lambda$ but the same formalism can be used to study periodic drives~\cite{HP2018}.

Due to the infinite range of interactions, assumed in the se\-cond term
of Eq.~(\ref{eq:HBCS}), the mean-field approximation is exact in the thermodynamic
limit. Hence, we consider the BCS mean-field Hamiltonian
which can be written in the Nambu basis as
%%%%%%%%%%%%%%%%%%%%%%%%%%%%
\begin{equation}
H_{\mathrm{MF}}=\sum_{\bm{k}}\psi_{\bm{k}}^{\dagger}\bm{H}_{\bm{k}}(t)\psi_{\bm{k}}^{}
\end{equation}
%%%%%%%%%%%%%%%%%%%%%%%%%%%
with $\psi_{\bm{k}}=\left(c_{\bm{k}\uparrow},c_{-\bm{k}\downarrow}^{\dagger}\right)^{T}$ and
%%%%%%%%%%%%%%%%%%%%%%%%%%%
\begin{equation}
\label{eq:hkt}
\bm{H}_{\bm{k}}(t)=\left(\begin{array}{cc}
\xi_{\bm{k}} & -\Delta(t)\\
-\Delta(t)^{*} & -\xi_{\bm{k}}
\end{array}\right)\,.
\end{equation}
%%%%%%%%%%%%%%%%%%%%%%%%%%%
The instantaneous superconducting order parameter is 
%%%%%%%%%%%%%%%%%%%%%%%%%%%
\begin{equation}\label{eq:ssd}
\Delta(t)=\lambda(t)\sum_{\bm{k}}\left\langle c_{\bm{k}\uparrow}^{\dagger}(t)c_{\bm{-k}\downarrow}^{\dagger}(t)\right\rangle 
\end{equation}
%%%%%%%%%%%%%%%%%%%%%%%%%%%
and $\left\langle \ldots\right\rangle $ denotes the expectation value on the initial state. 

The time-dependent perturbation, via a quench in the coupling constant $\lambda(t)$, injects energy into the system that will never dissipate if we only consider an isolated superconductor.
To describe dissipation effects the 
self-consistent solution of the gap equation is written in terms of the Keldysh two-time contour Green\textquoteright{}s functions which explicitly incorporates the
coupling with the environment.

%%%%%%%%%%%%%%%%%%%%%%%%%%%%%%%%%%%%%%%%%%%%%%%%%%%%%%%%%%%%%%%%%%%%%%%
\subsection{The out of equilibrium Green\textquoteright{}s functions}
%%%%%%%%%%%%%%%%%%%%%%%%%%%%%%%%%%%%%%%%%%%%%%%%%%%%%%%%%%%%%%%%%%%%%%%
The calculation is formulated in terms of the Keldysh two-time contour Green\textquoteright{}s
functions which in the Nambu spinor basis are $2\times 2$ matrices with matrix elements given by:
%%%%%%%%%%%%%%%%%%%%%%%%%%%
\begin{eqnarray}
\notag
\bm{G}_{\bm{k}}^{R}\left(t,t^{\prime}\right)_{\alpha\beta}&=&-i\theta\left(t-t^{\prime}\right)\left\langle \left\{ \psi_{\bm{k}{\alpha}}(t),\psi_{\bm{k}{\beta}}^{\dagger}\left(t^{\prime}\right)\right\} \right\rangle,\\
\notag
\bm{G}_{\bm{k}}^{A}\left(t,t^{\prime}\right)_{\alpha\beta}&=&i\theta\left(t^{\prime}-t\right)\left\langle \left\{ \psi_{\bm{k}{\alpha}}(t),\psi_{\bm{k}{\beta}}^{\dagger}\left(t^{\prime}\right)\right\} \right\rangle,\\
%\notag
%G_{\bm{k}}^{K}\left(t,t^{\prime}\right)_{\alpha\beta}&=&-i\left\langle \left[\psi_{\bm{k}{\alpha}}(t),\psi_{\bm{k}{\beta}}^{\dagger}\left(t^{\prime}\right)\right]\right\rangle,\\ 
\bm{G}_{\bm{k}}^{<}\left(t,t^{\prime}\right)_{\alpha\beta}&=&i\left\langle \psi_{\bm{k}{\alpha}}^{\dagger}\left(t^{\prime}\right)\psi_{\bm{k}{\beta}}(t)\right\rangle, 
\end{eqnarray}
%%%%%%%%%%%%%%%%%%%%%%%%%%%
where $R$, $A$ and $<$ correspond to the retarded, advanced and
lesser Green\textquoteright{}s functions, respectively. 
Notice that 
${\bm{G}}_{\bm{\bm{k}}}^{A }\left(t,t^{\prime}\right)={\bm{G}}_{\bm{\bm{k}}}^{R}\left(t^{\prime},t\right)^{\dagger}$, 
so only one of ${\bm{G}}_{\bm{\bm{k}}}^{A /R}$ needs to be computed. 
With these definitions the self-consistent Eq.~(\ref{eq:ssd}) for the time-dependent order parameter becomes, 
%%%%%%%%%%%%%%%%%%%%%%%%%%%
\begin{equation} 
\label{eq:deltadef}
\Delta(t)=-i \lambda(t) \sum_{\bm{k}} {\bm{G}_{\bm{k}}^{<}\left(t,t\right)}_{12}\,. 
\end{equation}
%%%%%%%%%%%%%%%%%%%%%%%%%%%

%
For a given time dependence of the order parameter $\Delta(t)$ (not necessarily self-consistent) and in the absence of coupling with the reservoir, the retarded and advanced Green functions are computed by solving the following differential equations (in matrix notation in the Nambu spinor basis and setting $\hbar=1$),
%%%%%%%%%%%%%%%%%%%%%%%%%%%
\begin{eqnarray}
\label{eq:retev}
\notag
{\bm{G}}_{\bm{\bm{k}}}^{R(0)}\left(t,t\right)&=&-i\bm{I},\\
%\notag
i\partial_{t}{\bm{G}}_{\bm{\bm{k}}}^{R(0)}\left(t,t^{\prime}\right)&=&\bm{H}_{\bm{k}}(t){\bm{G}}_{\bm{\bm{k}}}^{R(0)}\left(t,t^{\prime}\right),\;\;\;\;\;\; t>t^{\prime},\\
\notag
i\partial_{t^{\prime}}{\bm{G}}_{\bm{\bm{k}}}^{R(0)}\left(t,t^{\prime}\right)&=-&{\bm{G}}_{\bm{\bm{k}}}^{R(0)}\left(t,t^{\prime}\right)\bm{H}_{\bm{k}}\left(t^{\prime}\right),\;\;\;\;\;\; t>t^{\prime}.
\end{eqnarray}
%%%%%%%%%%%%%%%%%%%%%%%%%%%

In order to include the dissipation effects we use the Keldysh equations with self-energies encoding the coupling to a reservoir. 
Following Refs.~\cite{Moore2019} and \cite{Millis2017}, we use a mechanism for dissipation that associates to each pair of states $\bm{ k}\!\uparrow, -\bm{ k}\!\downarrow  $ its own
reservoir. The effect of the bath on the retarded Green function is dictated by Dyson equation \cite{Antipekka1994,Horacio1992},
\begin{eqnarray}
\label{eq:gl}
{\bm{G}}_{\bm{k}}^{R}(t,t^{\prime})&=&{\bm{G}}_{\bm{k}}^{R(0)}(t,t^{\prime}) \\
&+&
\int dt_{1}\int dt_{2}\,{\bm{G}}_{\bm{k}}^{R(0)}(t,t_{1}){\bm{\Sigma}}_{\bm{k}}^{R}(t_{1},t_{2}){\bm{G}}_{\bm{k}}^{R}(t_{2},t^{\prime}).\nonumber
\end{eqnarray} 
In the limit of a wide-band reservoir with identical coupling for each $\bm{k}$ (see Appendix~\ref{app:ap1}), 
 the retarded self-energy becomes $\bm{k}$-independent and assuming time translational invariance in the bath, the solution of the Dyson equation in the 
time domain results
%%%%%%%%%%%%%%%%%%%%%%%%%%%
\begin{equation}
\label{eq:retd}
{\bm{G}}_{\bm{k}}^{R}(t,t^{\prime})={\bm{G}}_{\bm{k}}^{R (0)}(t,t^{\prime})e^{-\gamma(t-t^{\prime})/2}\,.
\end{equation}
%%%%%%%%%%%%%%%%%%%%%%%%%%%
The parameter $\gamma$ describes the effects of inelastic scattering producing a finite lifetime $\tau=1/\gamma$  and a level broadening.

We emphasize that these equations are valid for an arbitrary time-dependence  $\Delta(t)$. 
To make the computation self-consistent one needs to use  $\Delta(t)$ from Eq.~(\ref{eq:deltadef}).
The lesser Green\textquoteright{}s function is given by  \cite{Antipekka1994,Horacio1992,Moore2019}
%%%%%%%%%%%%%%%%%%%%%%%%%%%
\begin{equation}
\label{eq:gl}
{\bm{G}}_{\bm{k}}^{<}(t,t^{\prime})=\int dt_{1}\int dt_{2}\,{\bm{G}}_{\bm{k}}^{R}(t,t_{1}){\bm{\Sigma}}_{\bm{k}}^{<}(t_{1},t_{2}){\bm{G}}_{\bm{k}}^{A}(t_{2},t^{\prime})\,,
\end{equation}
%%%%%%%%%%%%%%%%%%%%%%%%%%%
where the lesser self-energy ${\bm{\Sigma}}_{\bm{k}}^{<}\left(t_{1},t_{2}\right)$ can be written as a diagonal matrix ${\bm{\Sigma}}_{\bm{k}}^{<}\left(t_{1},t_{2}\right)={\bm{I}} \Sigma^{<}\left(t_{1},t_{2}\right) $ with
%%%%%%%%%%%%%%%%%%%%%%%%%%%
\begin{eqnarray}
\nonumber
\Sigma^{<}\left(t_{1},t_{2}\right)&=&i\gamma\int\frac{d\omega}{2\pi}f\left(\omega\right)e^{-i\omega\left(t_{1}-t_{2}\right)}\,\\
\label{eq:selfl}
&=&-\frac{1}{2\pi}\frac{\gamma}{t_{1}-t_{2}+i0^{+}}\,.
\end{eqnarray}
%%%%%%%%%%%%%%%%%%%%%%%%%%%
Here $f(\omega)$ is the Fermi function evaluated at the bath temperature and the last equality stands for the zero temperature limit (see Appendix~\ref{app:ap1} for details).
%%%%%%%%%%%%%%%%%%%%%%%%%%%%%%%%%%%%%%%%%%%%%
\subsection{The equilibrium state}
%%%%%%%%%%%%%%%%%%%%%%%%%%%%%%%%%%%%%%%%%%%%%

For a time-independent BCS Hamiltonian [Eq.~(\ref{eq:HBCS}) with a time-independent pairing interaction $\lambda$] the present formalism allows to study the effect of inelastic scattering at equilibrium. 
Since this is a source of pair breaking, the coupling to the bath has some important consequences that are known but usually obtained with different methods and which we recover here with Keldysh Green functions.

In equilibrium the time invariance is preserved and the retarded Green function with dissipation is given by Eq.~(\ref{eq:retd})
where ${\bm{G}}_{\bm{k}}^{R(0)}\left(t,t^{\prime}\right) \equiv {\bm{G}}_{ \bm{k}}^{R(0)}\left(t-t^{\prime}\right)$ only depends on $t-t^{\prime}.$
Thus, in the frequency domain the poles are shifted away from the real axis with an imaginary component $-i \gamma$ describing the levels broadening.
 
In the zero temperature limit, the self-consistency (c.f. Eq.~(\ref{eq:deltadef})) for the superconducting order parameter becomes
%%%%%%%%%%%%%%%%%%%%%%%%%%%
\begin{equation}
\label{eq:gape}
1=\frac{1}{\pi}\sum_{\bm{k}}\frac{\lambda}{E_{\bm{k}}}\arctan\left(\frac{2E_{\bm{k}}}{\gamma}\right)\,,
\end{equation}
%%%%%%%%%%%%%%%%%%%%%%%%%%%
where $E_{\bm{k}}=\sqrt{\xi_{\bm{k}}^2+\Delta ^2}$ is the undressed excitation energy (see Appendix~\ref{app:ap2} for details). As can be deduced from Eq.~(\ref{eq:gape}), the inelastic scattering 
reduces the equilibrium value of the superconducting order parameter and the critical temperature $T_{c}$ is given by the well-known expression 
$\ln(T_{c} / T^{0}_{c})=\psi(1/2+\gamma/T_{c})-\psi(1/2)$ where $T^{0}_{c}$ is the critical temperature for $\gamma=0$ and $\psi(x)$ is the Digamma function \cite{DeGennes}.

Finally, it is worth mentioning that the level broadening introduced here, leads to the widely used phenomenological density of states $\rho(\omega)$ for tunneling experiments that incorporates a Dynes parameter $\gamma$,  
%%%%%%%%%%%%%%%%%%%%%%%%%%%
\begin{equation}
\label{eq:dos}
\rho(\omega)=\rho_{0}\, \mathrm{Re} \left [ \frac{\omega+i \gamma}{\sqrt{(\omega+i \gamma)^2-\Delta^2}}\right ]\,,
\end{equation}
%%%%%%%%%%%%%%%%%%%%%%%%%%%
where $\rho_{0}$ is the normal phase density of states \cite{Hlubina2016,Hlubina2018}.
%%%%%%%%%%%%%%%%%%%%%%%%%%%%%%%%%%%%%%%%%%%%%%
\subsection{Out of equilibrium dynamics}
%%%%%%%%%%%%%%%%%%%%%%%%%%%%%%%%%%%%%%%%%%%%%%
The equal time lesser Green function at zero temperature has to be computed using Eqs.~(\ref{eq:retd}) and~(\ref{eq:gl}) leading  to
%%%%%%%%%%%%%%%%%%%%%%%%%%%%
\begin{eqnarray}
\label{eq:glf}
\nonumber
{\bm{G}}_{\bm{k}}^{<}(t)=-\frac{\gamma e^{-\gamma t}}{2\pi}\int_{-\infty}^{t}dt_{1}\int_{-\infty}^{t}dt_{2}\,&&{\bm{ G}}_{ \bm{\bm{k}}}^{R(0)}\left(t,t_{1}\right){\bm{ G}}_{\bm{k}}^{A(0)}(t_{2},t)\\
&&\times\frac{e^{\gamma(t_{1}+t_{2})/2}}{t_{1}-t_{2}+i0^{+}}  \,,
\end{eqnarray}
%%%%%%%%%%%%%%%%%%%%%%%%%%%%
where we have shortened the notation using a single time variable as the argument of ${\bm{G}}_{\bm{k}}^{<}$. For the sake of computational efficiency it is better not to use the integral form Eq.~(\ref{eq:glf}) but calculate the time derivative of ${\bm{G}}_{\bm{k}}^{<}$,
%%%%%%%%%%%%%%%%%%%%%%%%%%%%
\begin{equation}
\label{eq:gld}
\partial_{t}{\bm{G}}_{\bm{k}}^{<}(t)=-\gamma {\bm{G}}_{\bm{k}}^{<}(t)+\bm{\mathcal{I}}_{\bm{k}}(t)-i\left[\bm{H}_{\bm{k}}(t),{\bm{G}}_{\bm{k}}^{<}(t)\right]\,,
\end{equation}
%%%%%%%%%%%%%%%%%%%%%%%%%%%%
where
%%%%%%%%%%%%%%%%%%%%%%%%%%%%
\begin{equation}\label{eq:ik}
\bm{\mathcal{I}}_{\bm{k}}(t)=\frac{i\gamma}{2\pi}\int_{-\infty}^{t}dt^{\prime}\left(\frac{{\bm{G}}_{ \bm{k}}^{R (0)}\left(t,t^{\prime}\right)}{t-t^{\prime}-i0^{+}}+\frac{{\bm{G}}_{ \bm{k}}^{R(0)}\left(t,t^{\prime}\right)^{\dagger}}{t-t^{\prime}+i0^{+}}\right)e^{-\frac{\gamma(t-t^{\prime})}{2}}.
\end{equation}
%%%%%%%%%%%%%%%%%%%%%%%%%%%%
In equilibrium, the second term on the right hand side of Eq.~(\ref{eq:gld}) exactly cancels the first one and as ${\bm{G}}_{\bm{k}}^{<}$ commutes with the 
Hamiltonian, the stationary state is recovered. The effect of the bath is to introduce memory in the system, so in order to solve for the Green function, instead of
a differential equation local in time (as Eqs.~(\ref{eq:retev}) are) one needs to solve an integrodifferential equation which depends on the past evolution through
Eq.~(\ref{eq:ik}).

%%%%%%%%%%%%%%%%%%%%%%%%%%%%%%%%%%%%%%%%%%%%%%%%%%%%%
\begin{figure}[tb]
\includegraphics[width=0.5\textwidth]{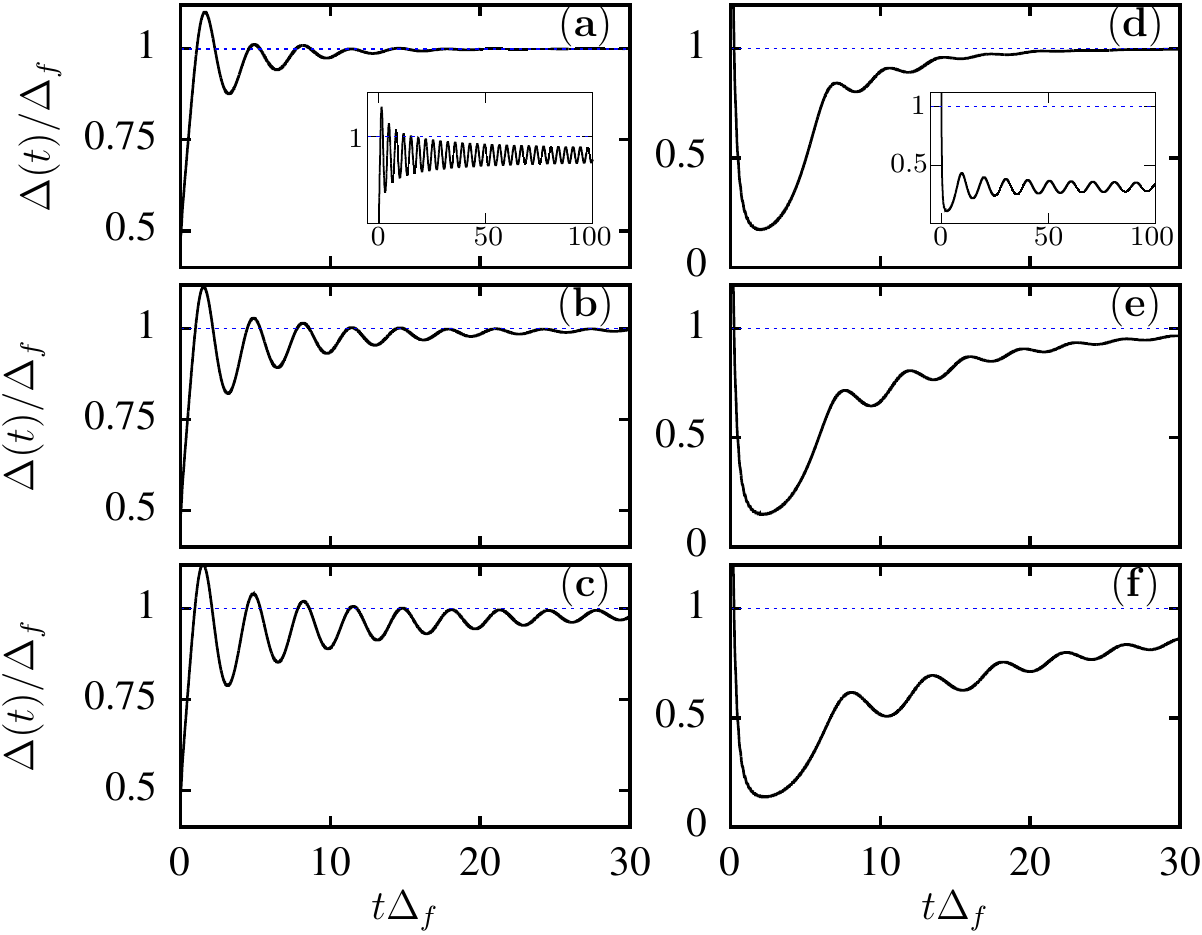}
\caption{(Color online) Time dependence of superconducting order parameter for $\Delta_{0}/\Delta_{f}=0.4$ (a-c) and $\Delta_{0}/\Delta_{f}=4$ (d-f). Upper, middle and bottom panel correspond with $\gamma=0.2\Delta_{f}$, $\gamma=0.1\Delta_{f}$ and
$\gamma=0.05\Delta_{f}$ respectively. In the inset of panel (a) and (d) we show $\Delta(t)$ in the dissipationless regime ($\gamma=0$) for $\Delta_{0}/\Delta_{f}=0.4$ and $\Delta_{0}/\Delta_{f}=4$ respectively (see Ref.\cite{Barankov2006a}).} 
\label{fig:fig1}
\end{figure}
%%%%%%%%%%%%%%%%%%%%%%%%%%%%%%%%%%%%%%%%%%%%%%%%%%%%%%%%

%%%%%%%%%%%%%%%%%%%%%%%%%%%%%%%%%%%
%\section{Results and Conclusions}
\section{Results}
%%%%%%%%%%%%%%%%%%%%%%%%%%%%%%%%%%

We now present results corresponding to a quench of the coupling parameter: $\lambda(t)=\theta(-t)\lambda_{0}+\theta(t)\lambda_{f}$. Eq.~(\ref{eq:gld}) is integrated
using a fourth order Runge-Kutta method with a small time steps in order to ensure the convergence of superconducting order parameter $\Delta (t)$. The lesser Green function
at equilibrium for $t<0$ is the initial condition for Eq.~(\ref{eq:gld}) (see Appendix~\ref{app:ap2}). To move forward in time $t$ we first integrate the third Eq.~(\ref{eq:retev}) in $t^{\prime}$ from $t$ to $t-10/\gamma$. This is used 
to construct the memory kernel Eqs.~(\ref{eq:ik}) at time $t$ needed to propagate forward in time the lesser Green function whith Eq.~(\ref{eq:gld}). In each time step the new value of $\Delta(t)$ is calculated and reinserted in the Hamiltonian of Eq.~(\ref{eq:hkt}). As we are interested in the low temperature regime, in order to optimize the computing time we used the zero temperature 
expression given in Eqs.~(\ref{eq:selfl}) and~(\ref{eq:glf}) and calculate the equilibrium value of $\Delta(t\rightarrow\infty)$ using Eq.~(\ref{eq:deltadef}).

In the following we parameterize the quantum quench not by the change of interaction constant but the ratio $\Delta_{0}/\Delta_{f}$, where $\Delta_{0}$ and $\Delta_{f}$ are the equilibrium  superconducting order parameters---satisfying the Eq.~(\ref{eq:gape})---for
$\lambda_{0}$ and $\lambda_{f}$, respectively. Note that a constant value of $\Delta_{0}/\Delta_{f}$ for different $\gamma$ values, implies different changes in $\lambda$.

Fig.~\ref{fig:fig1} shows the superconducting res\-pon\-se for moderate values of the quench parameter $\Delta_{0}/\Delta_{f}$, corresponding to the dephasing phase, and different values of $\gamma$. Panels (a), (b) and (c) correspond to an   increase of the coupling constant 
($\Delta_{0}/\Delta_{f}=0.4$ ) with $\gamma$ decreasing from top to bottom. In the absence of dissipation the order parameter is known~\cite{Barankov2004,Barankov2006a,Barankov2007} 
to oscillate with the Higgs-mode frequency $2 \Delta_{\infty}$ and
stabilize at long times at a value $\Delta_{\infty}<\Delta_{f}$. This is shown in the inset of panel (a). The effect of the bath is ({\em i}) to damp the 
oscillations and {(\em ii}) to introduce a slow drift so that $\Delta_{\infty}$ is replaced by the $T=0$ equilibrium value $\Delta_{f}$. 

In panels (d), (e) and (f) the results for a decrease of the coupling constant are shown. The order parameter decreases rapidly at short times and ``bounces back'' 
leading to the Higgs oscillations. Also in this case the asymptotic value in the absence of dissipation is $\Delta_{\infty}<\Delta_{f}$ as shown in the inset of panel (d). Again, the effect of the bath is to damp the oscillations and to 
introduce a drift toward  $\Delta_{f}$ with a time scale that becomes slower as $\gamma$ is decreased. 
.

%%%%%%%%%%%%%%%%%%%%%%%%%%%%%%%%%%%%%%%%%%%%%%%%%%%%%%%%
\begin{figure}[tb]
\includegraphics[width=0.5\textwidth]{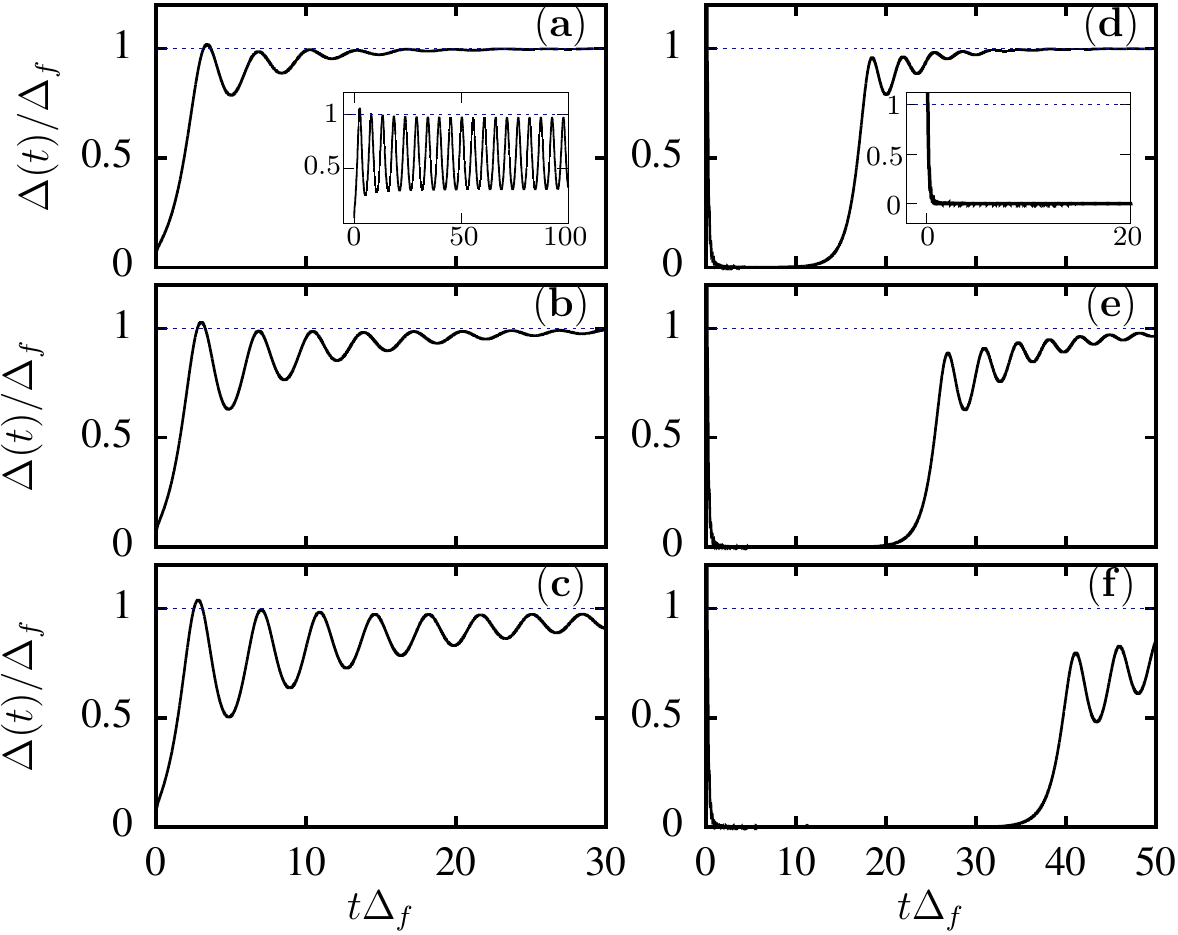}
\caption{(Color online)  Time dependence of superconducting order parameter for $\Delta_{0}/\Delta_{f}=0.05$ (a-c) and $\Delta_{0}/\Delta_{f}=8$ (d-f). Upper, middle and bottom panel correspond with $\gamma=0.2\Delta_{f}$, $\gamma=0.1\Delta_{f}$ and
$\gamma=0.05\Delta_{f}$ respectively. In the inset of panel (a) and (d) we show $\Delta(t)$ in the dissipationless regime ($\gamma=0$) for $\Delta_{0}/\Delta_{f}=0.05$ and $\Delta_{0}/\Delta_{f}=8$ respectively (see Ref.\cite{Barankov2006a}).} 
\label{fig:fig2}
\end{figure}
%%%%%%%%%%%%%%%%%%%%%%%%%%%%%%%%%%%%%%%%%%%%%%%%%%%%%%%%%%

The behavior for large quenches are shown in Fig.~\ref{fig:fig2}. In panels (a), (b) and (c) the order parameter increases driving the system to the synchronic regime (phase III) when $\gamma=0$ as shown in the inset of panel (a). 
To make the simulations affordable the characteristic time $1/\gamma$ was chosen of the same order of the simulation window ($t\Delta_f <30$) or smaller. For these parameters, thermalization takes place at times such that the 
synchronic (observed at long times in the disipationless case) and the dephasing  phases can hardly be distinguished. 

Panels (d), (e) and (f) of Fig.~\ref{fig:fig2} show the results for a large decrease of the coupling constant corresponding to the overdamped situation for the isolated system (phase II). After the quench, the order parameter decreases 
to an exponentially small value and remains small during a time interval which is controlled by the parameter $\gamma$. During this time interval the system thermalizes transferring energy to the bath without any noticeable effect. Remarkably, at some point the number of excitations 
becomes small enough and a fast increase of the superconducting order parameter is observed. From there on, the oscillatory evolution of $\Delta (t)$ towards its asymptotic value $\Delta_{f}$ takes place in which the amplitude of oscillations decay at a rate $e^{-\gamma t}$.

\begin{figure}[tb]
\includegraphics[width=0.5\textwidth]{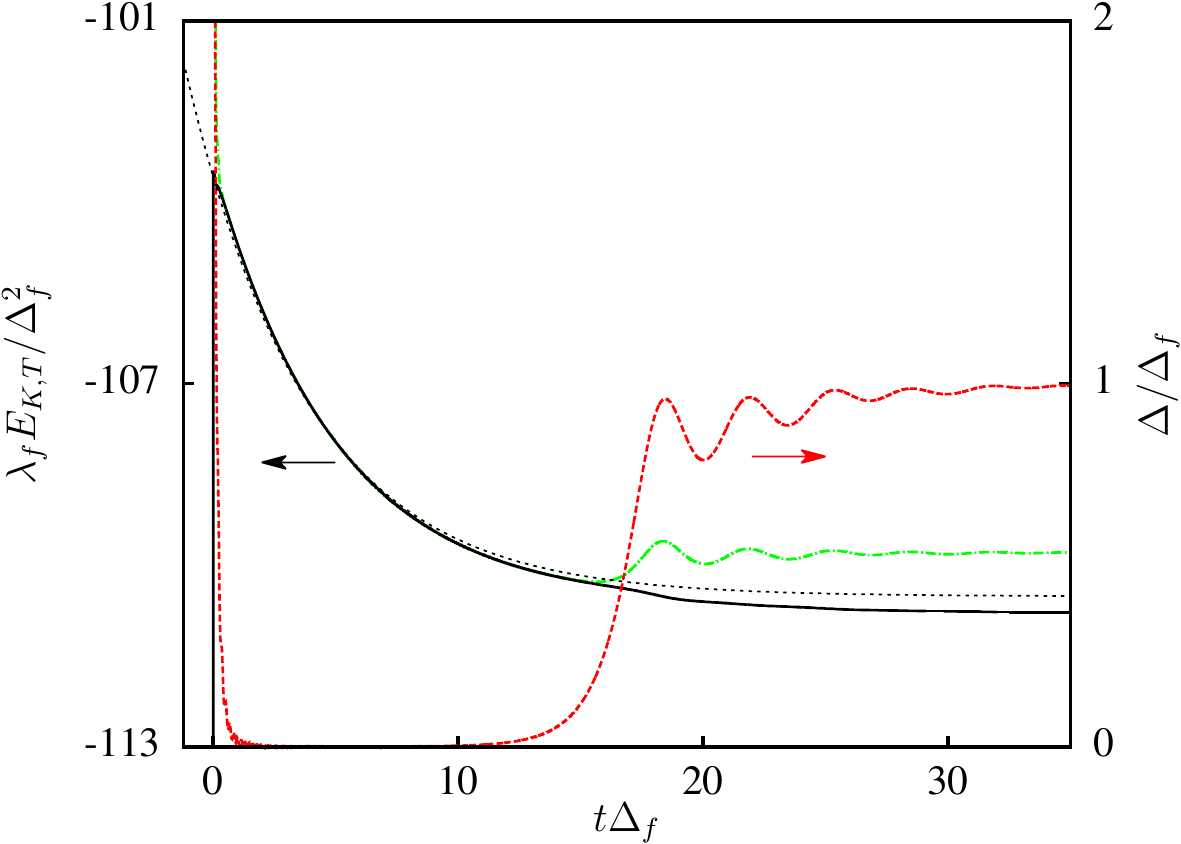}
\caption{(Color online) Kinetic energy (dash-dotted line, green on line), superconducting gap (dashed line, red online) and total energy (solid line) as a function of time for $\Delta_{0}/\Delta_{f}=8$ and 
$\gamma=0.2\Delta_{f}$ as in Fig.~\ref{fig:fig2} d).
The dotted line is the function $7 e^{-\gamma t}-110.5$ highlighting the exponential behavior of $E_{T}\sim E_{K}$ in an intermediate region where $\Delta(t)\sim0$.} 
\label{fig:fig3}
\end{figure}

To get more insight on this behavior we studied the 
total energy $E_{T}=\left\langle H_{MF} \right\rangle$ and the kinetic energy $E_{K}$ as a function of time. In terms of the lesser Green function
$E_{K}(t)= \sum_{\bm{k},\sigma}\xi_{\bm{k}}n_{\bm{k}\sigma}(t)=-i\sum_{\bm{k}}\xi_{\bm{k}}\left[\bm{G}_{\bm{k}}^{<}(t)_{11}-\bm{G}_{\bm{k}}^{<}(t)_{22}\right]$ where
$n_{\bm{k}\sigma}$ is the expectation value of the number operator. On the other hand, the interaction energy in the mean-field approximation is given by $E_{i}(t)=-\Delta(t)^2/\lambda(t)$ and 
$E_{T}(t)=E_{K}(t)+E_{i}(t).$
Fig.~\ref{fig:fig3} compare the evolution of kinetic and total energy with 
$\Delta (t)$  for the parameters of Fig.~\ref{fig:fig2}(d). Notice that at very short times after the quench ($t\Delta_{f}<0.5$)  the 
kinetic energy is larger than the total energy indicating a residual interaction energy $E_{i}(t)$.
In this first short transient $E_{K}(t)$ and $\Delta(t)$ decrease exponentially and $\Delta(t)$ goes to zero on the scale of the figure while the interaction energy (not shown) approaches zero from below. At $t\Delta_f\sim1$ the net effect of the quench 
is an excess of total energy constituted primarily of kinetic energy. This is because Cooper pairs are still rather localized but with random phases so they do not contribute to superconductivity.  As time evolves the excess kinetic energy is dissipated to the bath  decreasing as $e^{-\gamma t}$. In this regime the system behaves as a collection of free electrons with an out-of-equilibrium (non-thermal) distribution. Around $t\Delta_f\sim 10$ coherent superconductivity sets in again 
and $E_{K}$ increases as a result of the condensation of Cooper pairs and  the total energy decreases displaying a shallow shoulder. The system becomes rapidly a fully gaped superconductor again which is energetically more favorable. For longer times the total energy evolves towards 
its final equilibrium value.   The whole process resembles very much heating by the quench followed by cooling by the bath, however, one should keep in mind that only when the superconductor attains equilibrium with the bath a temperature can be defined (zero in our case).

%%%%%%%%%%%%%%%%%%%%%%%%%%%%%%%%%%%%%%%%%%%%%%%%%%%%%%%%%
%%%%%%%%%%%%%%%%%%%%%%%%%%%%%%%%%%%%%%%%%%%%%%%%%%%%%%%%%%

\section{ Conclusions}

In summary, we have revisited the problem of superconducting quenches 
incorporating the effect of the environment through a thermal bath at $T=0$.  
In all cases the quench implies an excess energy in the system respect to the final ground state. In the absence of dissipation this energy remains ``stored'' in the system and the order parameter reaches a stationary value 
smaller than the equilibrium value or show persistent oscillations. Clearly the effect of the bath is to absorb the excess energy driving the system to an equilibrium state.

Our results were obtained using a $\bm{k}$-independent relaxation time, an approximation that could be justified considering that all processes leading to the out
of equilibrium dynamics take place within a small energy window around the Fermi energy.  Nevertheless an extension to include a $\bm{k}$-dependent relaxation is 
straightforward: both, the equilibrium value $\Delta_{f}$ and out of equilibrium dynamics are obtained replacing $\gamma$ by $\gamma_{\bm{k}}$ in all expressions.

For $t\gg\tau=1/\gamma$ the superconducting order parameter reaches its thermal equilibrium irrespective of the strength of the quench  $\Delta(t\rightarrow\infty)=\Delta_{f}$, as expected. 
As mentioned above, in dissipationless weak coupling BCS systems, a small quench excites the Higgs mode that, due to the dephasing, decays with a power law $t^{-1/2}$ \cite{Matsunaga2013}. In contrast, in the strong coupling limit, the exponent
increases to reach the value $3/2$ \cite{Gurarie2009}. It has also been shown that for isolated systems in the strong coupling limit with non-local pairing interaction, the synchronic phase
is much more stable and the Higgs mode becomes undamped both for an increase and a decrease of the coupling parameter \cite{Barankov2007}. However, all these asymptotic properties manifest at long times. Therefore, even for small values of $\gamma$ the exponential decay may dominate, making very challenging  to conclude on the dephasing time exponent or the undamped character of the Higgs mode in condensed matter systems. Fermionic cold atoms, 
with a high degree of coherence and tunable interactions, may offer new opportunities for the experimental study of the physics described above.

The heart of quantum information processing is to exploit non-linear effects that appear when coupled superconductors elements are excited by external drives far from equilibrium states~\cite{wending2018}. Of paramount importance is 
to gauge the role of the environment on these manipulations. Our computations set a framework for studying such dynamics which will play a key role on the future design of quantum technological devices.

%%%%%%%%%%%%%%%%%%%%%%%%%%%%%%%%%%%%%%%%%%%%%%55555
\begin{acknowledgments}
 We acknowledge financial support 
from Italian MAECI and Argentinian MINCYT through  bilateral project AR17MO7 
and  Italian MAECI thought collaborative project SUPERTOP-PGR04879.
We acknowledge financial support from ANPCyT (grant PICT 2016-0791),  CONICET (grant PIP 11220150100506) and SeCyT-UNCuyo (grant 06/C526)
 and from Regione Lazio (L.R. 13/08) under project SIMAP. 

\end{acknowledgments}

%%%%%%%%%%%%%%%%%%%%%%%%%%%%%%%%%%%%%%%%%%%%%%
\appendix
%%%%%%%%%%%%%%%%%%%%%%%%%%%%%%%%%%%%%%%%5
\section{Lesser self-energy to include dissipation}
\label{app:ap1}
In our computations to each pair of states ${\bm k} \uparrow, -{\bm k} \downarrow  $ we associate a heat bath described by a time-independent free-particle 
Hamiltonian $H_{b}=\sum_{\ell,\sigma}E_{\ell}d_{\ell\sigma}^{\dagger}d_{\ell\sigma}$, where
$d_{\ell\sigma}^{\dagger}$ creates an electron in state $\left|\ell\right\rangle$ of the reservoir with energy $E_{\ell}$ and spin $\sigma$.
Thus, the time translational invariance in the bath allows us to deal with self-energies in the frequency space. In the Nambu basis $H_{b}$ takes the form of Eq.~(\ref{eq:hkt}) with $\Delta(t)=0$ and the retarded self-energy 
$\mathbf{\Sigma}^{R}_{\bm k}$ can be written as
%%%%%%%%%%%%%%%%%%%%%%%%%%%%
\begin{flalign}
\mathbf{\Sigma}^{R}_{\bm{k}}\left(\omega\right)&=\sum_{\ell}\left|V_{\bm{k}\ell}\right|^{2}g_{\ell}^{R}\left(\omega\right)&&\\\nonumber
&=\sum_{\ell}\left|V_{\bm{k}\ell}\right|^{2}\left(\begin{array}{cc}
\left(\omega-E_{\ell}+i0^{+}\right)^{-1} & 0\\
0 & \left(\omega+E_{\ell}+i0^{+}\right)^{-1}
\end{array}\right)
\end{flalign}
%%%%%%%%%%%%%%%%%%%%%%%%%%%%
where $V_{\bm{k}\ell}$ is the coupling between superconducting quasiparticles and
reservoir and $g_{\ell}^{R}\left(\omega\right)$ is the retarded
Green\textquoteright{}s function corresponding with $H_{b}.$ For simplicity, in the following, we drop the ${\bm k}$ label from all quantities by assuming a 
$\bm{k}-$independent coupling $V_{\bm{k}\ell}\equiv V_{\ell}$. 

Therefore the nonzero matrix elements of retarded self-energy read, 
%%%%%%%%%%%%%%%%%%%%%%%%%%%%
\begin{equation}
\bm{\Sigma}^{R}\left(\omega\right)_{11}=\sum_{\ell}\frac{\left|V_{l}\right|^{2}}{\omega-E_{\ell}+i0^{+}}=\Lambda\left(\omega\right)-\frac{i}{2}\gamma\left(\omega\right)
\end{equation}
%%%%%%%%%%%%%%%%%%%%%%%%%%%%
and $\bm{\Sigma}^{R}\left(\omega\right)_{22}=-\bm{\Sigma}^{R}\left(-\omega\right)^{*}_{11}$.
In the following, we use the wideband approximation to neglect $\Lambda\left(\omega\right)$
and assume that the level broadening $\gamma$ is an energy-independent
parameter of our model, which defines the time scale for dissipation.
As a consequence, 
\begin{eqnarray}
  \label{eq:gamma}
\bm{\Sigma}^{R}\left(\omega\right)=-i\bm{I}\gamma/2,
\end{eqnarray}
and  the solution of the Dyson equation for retarded Green function is given by Eq.~(\ref{eq:retd}).
In another hand, the lesser self-energy is
%%%%%%%%%%%%%%%%%%%%%%%%%%%%
\begin{equation}
\bm{\Sigma}^{<}\left(\omega\right)=\sum_{\ell}\left|V_{\ell}\right|^{2}g_{\ell}^{<}\left(\omega\right)=i\gamma f\left(\omega\right)\bm{I}\,,
\end{equation}
%%%%%%%%%%%%%%%%%%%%%%%%%%%%
where $g_{\ell}^{<}\left(\omega\right)=if(\omega)A_{\ell}(\omega)$ is the lesser Green\textquoteright{}s
function associated to $H_{b}$, $A_{\ell}(\omega)$ is the spectral function and $f\left(\omega\right)=\theta\left(-\omega\right)$
is the Fermi function at zero temperature. The self-energy in time domain can then be written as
%%%%%%%%%%%%%%%%%%%%%%%%%%%%
\begin{equation}
\label{eq:selfa}
\bm{\Sigma}^{<}\left(t_{1}-t_{2}\right)=i\gamma\bm{I}\int\frac{d\omega}{2\pi}\theta\left(-\omega\right)e^{-i\omega\left(t_{1}-t_{2}\right)}=\frac{-\gamma/2\pi}{t_{1}-t_{2}+i0^{+}}\bm{I}.
\end{equation}
%%%%%%%%%%%%%%%%%%%%%%%%%%%%
%%%%%%%%%%%%%%%%%%%%%%%%%%%%%%%%%%%%%%%%%%%%%%%%%%%%%%%%%%%%%%%%%%%%%%%%%
\section{Lesser Green function and gap equation at equilibrium }
\label{app:ap2}
%%%%%%%%%%%%%%%%%%%%%%%%%%%%%%%%%%%%%%%%%%%%%%%%%%%%%%%%%%%%%%%%%%%%%%
In this section we present derivation of equilibrium lesser Green function from the proposed self-energy Eq.~(\ref{eq:selfa}) by using Eq.~(\ref{eq:glf}). 
A similar procedure was carried out in the appendix of Ref \cite{Moore2019} where approximate expressions for the trivial $\gamma\rightarrow  0$ limit were used while here we evaluate the integral exactly for all $\gamma$ values.
The equilibrium expressions of retarded and advanced Green functions are given by
\begin{widetext}
%%%%%%%%%%%%%%%%%%%%%%%%%%%%
\begin{equation}
\bm{G}_{\bm{\bm{k}}}^{R}\left(t,t^{\prime}\right)=-i\theta\left(t-t^{\prime}\right)\left(\begin{array}{cc}
u_{\bm{k}}^{2}e^{-iE_{\bm{k}}\left(t-t^{\prime}\right)}+v_{\bm{k}}^{2}e^{iE_{\bm{k}}\left(t-t^{\prime}\right)} & -u_{\bm{k}}v_{\bm{k}}\left(e^{-iE_{\bm{k}}\left(t-t^{\prime}\right)}-e^{iE_{\bm{k}}\left(t-t^{\prime}\right)}\right)\\
-u_{\bm{k}}v_{\bm{k}}\left(e^{-iE_{\bm{k}}\left(t-t^{\prime}\right)}-e^{iE_{\bm{k}}\left(t-t^{\prime}\right)}\right) & u_{\bm{k}}^{2}e^{iE_{\bm{k}}\left(t-t^{\prime}\right)}+v_{\bm{k}}^{2}e^{-iE_{\bm{k}}\left(t-t^{\prime}\right)}
\end{array}\right)e^{-\gamma(t-t^{\prime})/2}
\end{equation}
%%%%%%%%%%%%%%%%%%%%%%%%%%%%
\end{widetext}
and $\bm{G}_{\bm{\bm{k}}}^{A}\left(t,t^{\prime}\right)=\bm{G}_{\bm{\bm{k}}}^{R}\left(t^{\prime},t\right)^{\dagger}$
where 
%%%%%%%%%%%%%%%%%%%%%%%%%%%
\begin{equation}
E_{\bm{k}}=\sqrt{\xi_{\bm{k}}^{2}+\Delta_{0}^{2}},\,\,\, u_{\bm{k}}^{2}=\frac{1}{2}\left(1+\frac{\xi_{\bm{k}}}{E_{\bm{k}}}\right),\,\,\,
v_{\bm{k}}^{2}=\frac{1}{2}\left(1-\frac{\xi_{\bm{k}}}{E_{\bm{k}}}\right)\\
\end{equation}
%%%%%%%%%%%%%%%%%%%%%%%%%%%%%%
and $\Delta_{0}$ is the order parameter before quantum quench which is set to be real without loss of generality. After
introduce these expressions in Eq.~(\ref{eq:glf}) the components of $\bm{G}_{\bm{k}}^{<}(t)$
are given by: 
%\begin{widetext}
%%%%%%%%%%%%%%%%%%%%%%%%%%%%
\begin{eqnarray}
\nonumber
\bm{G}_{\bm{k}}^{<}(t)_{11}&=&-\frac{\gamma e^{-\gamma t}}{2\pi}\int_{-\infty}^{t}dt_{1}\int_{-\infty}^{t}dt_{2}\frac{e^{\gamma(t_{1}+t_{2})/2}}{t_{1}-t_{2}+i0^{+}} \\ 
&&\times \left(u_{\bm{k}}^{2}e^{iE_{\bm{k}}\left(t_{1}-t_{2}\right)}+v_{\bm{k}}^{2}e^{-iE_{\bm{k}}\left(t_{1}-t_{2}\right)}\right)\,,\\
\nonumber
\bm{G}_{\bm{k}}^{<}(t)_{12}&=&\frac{i\gamma u_{\bm{k}}v_{\bm{k}}e^{-\gamma t}}{2\pi}\int_{-\infty}^{t}dt_{1}\int_{-\infty}^{t}dt_{2}\frac{e^{\gamma(t_{1}+t_{2})/2}}{t_{1}-t_{2}+i0^{+}} \\
&& \times \sin(E_{\bm{k}}\left(t_{1}-t_{2}\right))\,,
\end{eqnarray}
%%%%%%%%%%%%%%%%%%%%%%%%%%%%
%\end{widetext}
$\bm{G}_{\bm{k}}^{<}(t)_{21}=\bm{G}_{\bm{k}}^{<}(t)_{12}$
and $\bm{G}_{\bm{k}}^{<}(t)_{22}=\bm{G}_{\bm{k}}^{<}(t)_{11}$
after the interchange $u_{\bm{k}}^{2}\leftrightarrow v_{\bm{k}}^{2}.$
By introducing the change of variables $T=(t_{1}+t_{2})/2$ and $\tau = t_{1}-t_{2}$, 
the Green functions read
%%%%%%%%%%%%%%%%%%%%%%%%%%%%
\begin{eqnarray}
\bm{G}_{\bm{k}}^{<}(t)_{11}&=&-\frac{\gamma e^{-\gamma t}}{2\pi}\int_{-\infty}^{t}dTe^{\gamma T}\times\\
\nonumber
&&\int_{-2(t-T)}^{2(t-T)}d\tau\left[\frac{\cos(E_{\bm{k}}\tau)}{\tau+i0^{+}}+ i\frac{(u_{\bm{k}}^{2}-v_{\bm{k}}^{2})\sin(E_{\bm{k}}\tau)}{\tau+i0^{+}}\right]\,
\end{eqnarray}
%%%%%%%%%%%%%%%%%%%%%%%%%%%%%
and
%%%%%%%%%%%%%%%%%%%%%%%%%%%%%
\begin{eqnarray}
\nonumber
\bm{G}_{\bm{k}}^{<}(t)_{12}&=&\frac{i\gamma u_{\bm{k}}v_{\bm{k}}e^{-\gamma t}}{\pi}\int_{-\infty}^{t}dTe^{\gamma T}\int_{-2(t-T)}^{2(t-T)}d\tau\frac{\sin(E_{\bm{k}}\tau)}{\tau+i0^{+}}\,.\\
\end{eqnarray}
%%%%%%%%%%%%%%%%%%%%%%%%%%%%
Thus, since
%%%%%%%%%%%%%%%%%%%%%%%%%%%%
\begin{equation}
\int_{-2(t-T)}^{2(t-T)}d\tau\frac{\sin(E_{\bm{k}}\tau)}{\tau+i0^{+}}=2\mathrm{Si}\left(2E_{\bm{k}}(t-T)\right) 
\end{equation}
%%%%%%%%%%%%%%%%%%%%%%%%%%%%
where $\mathrm{Si}$ represents the sine integral, and 
%%%%%%%%%%%%%%%%%%%%%%%%%%%%
\begin{equation}
\int_{-2(t-T)}^{2(t-T)}d\tau\frac{\cos(E_{\bm{k}}\tau)}{\tau+i0^{+}}=-i\pi, 
\end{equation}
%%%%%%%%%%%%%%%%%%%%%%%%%%%%
we can write
%%%%%%%%%%%%%%%%%%%%%%%%%%%% 
\begin{equation}
\bm{G}_{\bm{k}}^{<}(t)_{11}=\frac{i}{2}-\frac{i\gamma\xi_{\bm{k}}e^{-\gamma t}}{\pi E_{\bm{k}}}\int_{-\infty}^{t}dTe^{\gamma T}\mathrm{Si}\left(2E_{\bm{k}}(t-T)\right)
\end{equation}
%%%%%%%%%%%%%%%%%%%%%%%%%%%%
and
\begin{equation}
\bm{G}_{\bm{k}}^{<}(t)_{12}=\frac{i\gamma\Delta_{0}e^{-\gamma t}}{\pi E_{\bm{k}}}\int_{-\infty}^{t}dTe^{\gamma T}\mathrm{Si}\left(2E_{\bm{k}}(t-T)\right).
\end{equation}
Finally, after compute these integrals we obtain the equilibrium lesser Green function
\begin{equation}
\bm{G}_{\bm{k}}^{<}(t)=\frac{i}{2}\bm{I}-\frac{i}{\pi E_{\bm{k}}}\arctan\left(\frac{2E_{\bm{k}}}{\gamma}\right)\left(\begin{array}{cc}
\xi_{\bm{k}} & -\Delta_{0}\\
-\Delta_{0} & -\xi_{\bm{k}}
\end{array}\right)
\end{equation}
which has been used as the initial condition for the differential Eq.~(\ref{eq:gld}) at $t=0$.
Therefore, the superconducting
gap $\Delta_{0}=-i\lambda_{0}\sum_{\bm{k}}{{\bm{G}_{\bm{k}}}^{<}(t)}_{12}$ is obtained via the gap equation 
~(\ref{eq:gape}) with $\lambda=\lambda_{0}$.


\begin{thebibliography}{10}

\bibitem{Stojchevska2014}
L. Stojchevska, I. Vaskivskyi, T. Mertelj, P. Kusar,
D. Svetin, S. Brazovskii, and D. Mihailovic, Science {\bf 344}, 177 (2014).

\bibitem{Mitrano2016}
M. Mitrano, A. Cantaluppi, D. Nicoletti, S. Kaiser, A. Perucchi,
S. Lupi, P. Di Pietro, D. Pontiroli, M. Ricc`o, S. R.
Clark, D. Jaksch, and A. Cavalleri, Nature {\bf 530}, 461 (2016).

\bibitem{Fausti2011}
D. Fausti, R. I. Tobey, N. Dean, S. Kaiser, A. Dienst, M. C. Hoffmann, S. Pyon, T. Takayama, H. Takagi, and A. Cavalleri, Science {\bf 331}, 189 (2011).

\bibitem{Matsunaga2013}
R. Matsunaga, Y.~I. Hamada, K. Makise, Y. Uzawa, H. Terai, Z. Wang, and R.
  Shimano, Phys. Rev. Lett. {\bf 111},  057002  (2013).
  
\bibitem{Matsunaga2014}
R. Matsunaga, N. Tsuji, H. Fujita, A. Sugioka, K. Makise, Y. Uzawa, H. Terai,
  Z. Wang, H. Aoki, and R. Shimano, Science {\bf 345},  1145  (2014).
    
\bibitem{Mansart2013}
B. Mansart, J. Lorenzana, a. Mann, a. Odeh, M. Scarongella, M. Chergui, and F.
  Carbone, Proc. Natl. Acad. Sci. {\bf 110},  4539  (2013).
  
\bibitem{Stefanucci}
 G. Stefanucci and R. van Leeuwen, Nonequilibrium ManyBody
Theory of Quantum Systems: A Modern Introduction
(Cambridge University Press, 2013).

\bibitem{Stan2009}
A. Stan, N. E. Dahlen and R. van Leeuwen, J. Chem. Phys. {\bf 130}, 224101 (2009).

\bibitem{Stefanucci2009}
P. Myöhänen, A. Stan, G. Stefanucci, R. van Leeuwen, Phys. Rev. B {\bf 80}, 115107 (2009).

\bibitem{Leeuwen2007}
Nils Erik Dahlen and Robert van Leeuwen, Phys. Rev. Lett. {\bf 98}, 153004 (2007).

\bibitem{Ecksteini2009}
M. Eckstein, M. Kollar, P. Werner, Phys. Rev. Lett. {\bf 103}, 056403 (2009).

\bibitem{Ecksteini2014}
H. Aoki, N. Tsuji, M. Eckstein, M. Kollar, T. Oka, P. Werner, Rev. Mod. Phys. {\bf 86}, 779 (2014).

\bibitem{Barankov2004}
R.~A. Barankov, L.~S. Levitov, and B.~Z. Spivak, Phys. Rev. Lett. {\bf 93},
  160401  (2004).

\bibitem{Barankov2006a}
R.~A. Barankov and L.~S. Levitov, Phys. Rev. Lett. {\bf 96},  230403  (2006).

\bibitem{Yuzbashyan2006a}
E.~A. Yuzbashyan and M. Dzero, Phys. Rev. Lett. {\bf 96},  230404  (2006).

\bibitem{Hannibal2018}
S. Hannibal, P. Kettmann, M.~D. Croitoru, V.~M. Axt, and T. Kuhn, Phys. Rev. A
  {\bf 97},  013619  (2018).

\bibitem{Hannibal2018a}
S. Hannibal, P. Kettmann, M.~D. Croitoru, V.~M. Axt, and T. Kuhn, Phys. Rev. A
  {\bf 98},  053605  (2018).

\bibitem{Yuzbashyan2005}
E.~A. Yuzbashyan, B.~L. Altshuler, V.~B. Kuznetsov, and V.~Z. Enolskii, J.
  Phys. A. Math. Gen. {\bf 38},  7831  (2005).

\bibitem{Yuzbashyan2006}
E.~A. Yuzbashyan, O. Tsyplyatyev, and B.~L. Altshuler, Phys. Rev. Lett. {\bf
  96},  097005  (2006).
  
\bibitem{HP2018}
H.~P. Ojeda Collado, Jos\'e Lorenzana, Gonzalo Usaj, and C. A. Balseiro,  Phys. Rev. B {\bf 98},  214519  (2018).



\bibitem{Cea2016}
T. Cea, C. Castellani, and L. Benfatto, Phys. Rev. B {\bf 93},  180507  (2016).

\bibitem{Antipekka1994}
Antti-Pekka Jauho, Ned S. Wingreen, and Yigal Meir Phys. Rev. B {\bf 50}, 5528 (1994)

\bibitem{Moore2019}
Tianrui Xu, Takahiro Morimoto, Alessandra Lanzara and Joel E. Moore,  Phys. Rev. B {\bf 99},  035117  (2019).

\bibitem{Horacio1992}
Horacio M. Pastawski,  Phys. Rev. B {\bf 46},  4053 (1992).

\bibitem{Millis2017}
D. M. Kennes, E. Y. Wilner, D. R. Reichman, and A. J. Millis,  Phys. Rev. B {\bf 96},  054506  (2017).

\bibitem{Gurarie2009}
V. Gurarie,  Phys. Rev. Lett. {\bf 103},  075301  (2009).

\bibitem{Barankov2007}
R.~A. Barankov and L.~S. Levitov,  arXiv:0704.1292.

\bibitem{DeGennes}
P. G. DeGennes, Superconductivity of Metals and Alloys (Addison-Wesley, Reading, MA, 1989).

\bibitem{Hlubina2016}
František Herman and Richard Hlubina,  Phys. Rev. B {\bf 94},  144508  (2016).

\bibitem{Hlubina2018}
František Herman and Richard Hlubina,  Phys. Rev. B {\bf 97},  014517  (2018).

\bibitem{wending2018}
G Wendin, Rep. Prog. Phys. {\bf 80}  106001 (2017).


\end{thebibliography}
\end{document}